# Locate the Source of Resonance-Involved Forced Oscillation in Power Systems Based on Mode Shape Analysis

Shutang You
University of Tennessee, Knoxville, TN, USA
Email: syou3@utk.edu

*Abstract*— This paper proposed a new method to locate the source of forced oscillation that involves resonance with natural oscillation modes. The new method is based on comparing the oscillation mode shape of the forced oscillation with that of the natural oscillation that the forced oscillation resonates with. The oscillation source is the location that has the largest angle difference between the forced oscillation mode and the natural oscillation mode. Some examples in the actual U.S. EI system verified this approach.

## I. Introduction

Power grid reliability is important to ensure continuous electricity supply. The recent increase of renewable generation is also changing the dynamics of power grids and consequently how the grid is operated, and how its reliability is maintained. Based on synchrophasor technology, wide-area monitoring systems (WAMS) have been developed to help improve the situational awareness of power grids. WAMS data have been proven useful in detecting and analyzing the oscillations in power systems.

Forced oscillations are usually caused by malfunctions of controls. They are typical shown as persistent oscillations in system with nearly zero damping. Since the location of the driving force is important information to mitigate forced oscillations, literature has been focusing on locating the source of forced oscillations.

Locating the source of forced oscillations is a challenging task, because in some cases, the location with the largest oscillation amplitude is not the actual oscillation source. This phenomenon is more common if the oscillation driving force has an oscillatory frequency close to the natural oscillation mode of a system, which often involves resonance [1]. To deal with this problem, some literature has made progresses. For example, Ref. [2] used the energy based method to locate the source of forced oscillations. Ref. [3] used a data-driven method to locate the force oscillation source. Ref. [4] provides a system identification technology to locate the oscillation source.

In this paper, the forced oscillation is located by comparing the oscillation mode shapes between the forced oscillation and the natural oscillation that the forced oscillation resonates with. Two case studies in the U.S. EI system is used to verify the proposed method.

## II. Locate the Source of Resonance-Involved Forced Oscillation in Power Systems Based on Mode Shape Analysis

For the particular type of forced oscillation that involves resonance, the driving force frequency is close to natural oscillation frequency. The forced oscillation mode and natural oscillation mode would also be very similar. Except for that the mode phase arrow near the driving force location will be distorted due to the driving force that consistently feeds energy into the natural oscillation. Therefore, the location of power system forced oscillation source includes the following steps:

1. Plot the oscillation mode shape based on the time-domain frequency data from multiple PMUs at different locations.
2. Set a threshold of the magnitude to select the PMU frequency data that have relatively large oscillation magnitudes.
3. The angles of the selected PMU frequency channels in the forced oscillation mode are shifted by a constant angle to maximize its fitness level with the natural oscillation mode shape that the forced oscillation resonates to. This can be done by minimizing an objective function of all mode shape angles' RMS value between the forced oscillation mode and natural resonance oscillation mode.
4. Check several largest values of the angle difference between the shifted forced oscillation mode shape and the natural oscillation mode. If one angle difference is significantly larger than the rest, then the location of this PMU data is the location of the oscillation source. If several locations have comparable angle differences among the largest, then the forced oscillation source can be obtained using a triangulation method based on the angle difference values [5].

## III. Case Studies in the U.S. Eastern Interconnection Power Grid

The proposed method is applied to two actual oscillation cases in the U.S. EI system, as shown in TABLE I.

This work made use of Engineering Research Center shared facilities supported by the Engineering Research Center Program of the National Science Foundation and the Department of Energy under NSF Award Number EEC-1041877 and the CURENT Industry Partnership Program.

TABLE I. Two forced oscillation cases in the U.S. EI system

| Case | Time | Oscillation source location | Oscillation frequency |
|---|---|---|---|
| 1 | 01/11/2019 | Florida | 0.25 Hz |
| 2 | 11/27/2016 | Georgia | 0.7 Hz |

Figure 1 shows a forced oscillation event happened on Jan 11 2019. This event lasted for around 18 minutes before the malfunction generator was removed from the system. Figure 2 shows the mode shape of this forced oscillation when the system has a driving force. Figure 3 shows the oscillation mode shape of the system natural oscillation mode that has a very close frequency as the forced oscillation. Using the proposed method, the two oscillation modes are first aligned together, as shown in Figure 4. It is clear that the largest angle value difference is in Florida. Therefore, it can be inferred that Florida is the oscillation source location. NERC also confirmed the driving source of this forced oscillation event is in Florida [6].

Figure 5 and Figure 6 show the oscillation and its mode shape when the driving force (i.e. the malfunction unit) was removed from the system. The mode angle of Florida returned to be consistent with the oscillation angle in the Southern Company (SoCo). This also confirmed that proposed method is effective is locating the oscillation source.

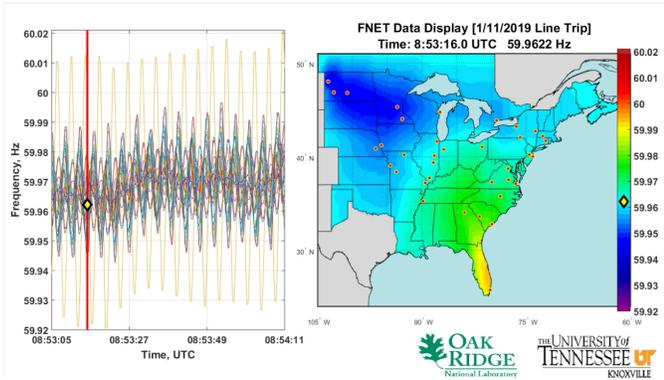
Figure 1. U.S. EI system oscillation on Jan. 11 2019. [7]

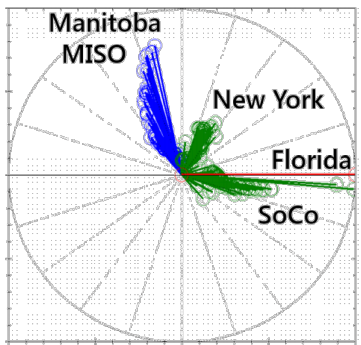
Figure 2. Oscillation mode of the forced oscillation at 0.25Hz.[6]

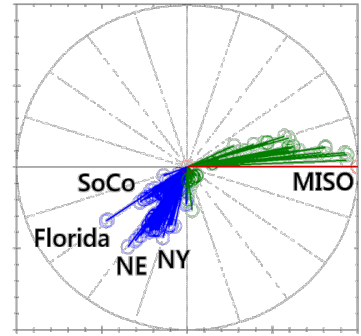
Figure 3. Oscillation mode of the natural oscillation at 0.25Hz.[6]

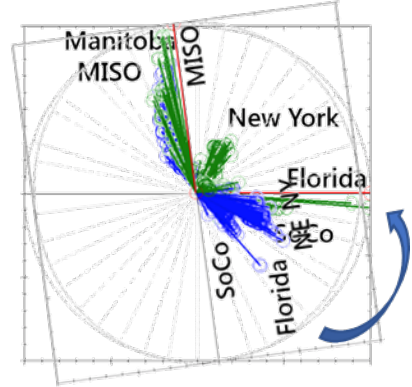
Figure 4. Mode angle difference check after aligning the forced oscillation and the natural oscillation modes (Jan 11 2019 event)

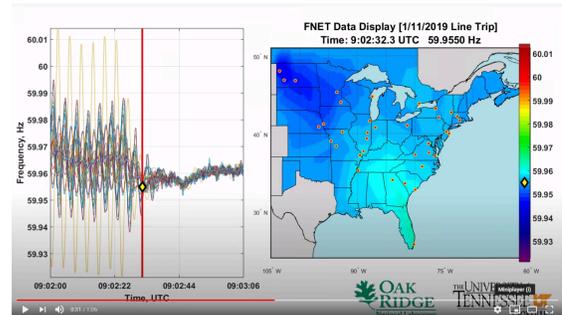
Figure 5. U.S. EI system oscillation after removing the forced oscillation source on Jan. 11 2019.[7]

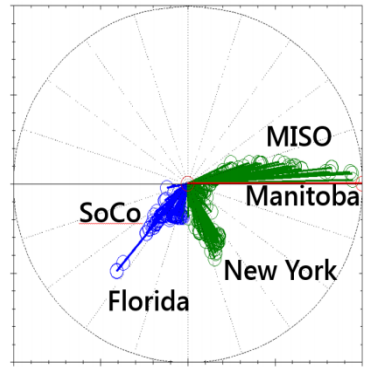
Figure 6. Oscillation identified during ring-down period after removing the oscillation driving force on Jan 11 2019.[6]

Figure 7 shows the forced oscillation event happened in Nov 27 2016. Figure 8 and Figure 9 show the mode shapes of the forced oscillation and the natural oscillation respectively. Both two oscillation modes have a frequency close to 0.7Hz. Aligning up the two oscillation modes, it can be seen that Georgia has the largest angle difference between the forced oscillation mode and the natural oscillation mode. The utility companies also confirmed that the source of this forced oscillation event is located in Georgia [8].

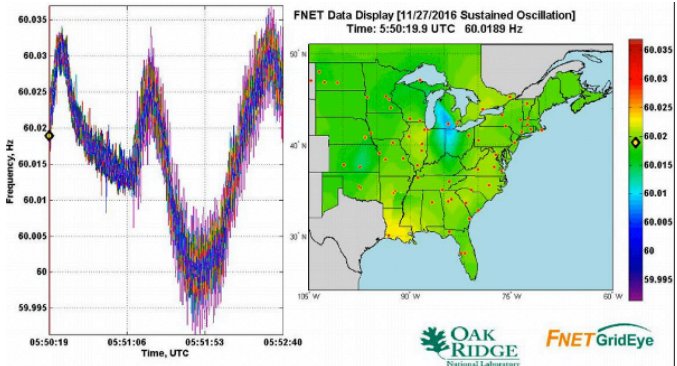

Figure 7. U.S. EI system oscillation on Nov. 27 2016.[7]

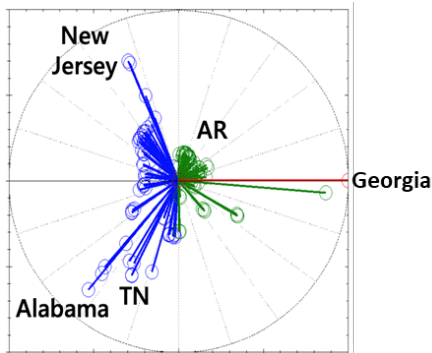

Figure 8. Oscillation mode of the forced oscillation at 0.7 Hz.[8, 9]

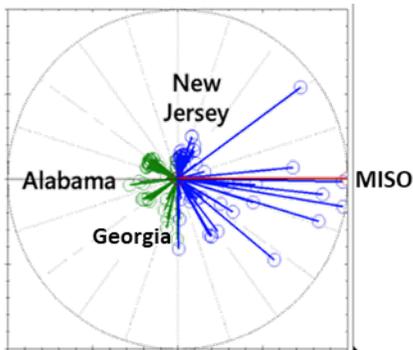

Figure 9. Oscillation mode of the natural oscillation at 0.75 Hz.[8, 9]

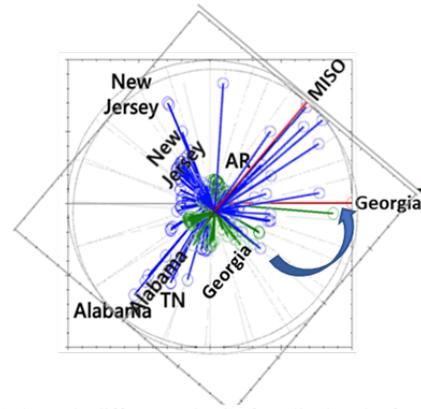

Figure 10. Mode angle difference check after aligning the forced oscillation and the natural oscillation modes (Nov. 27 2016 event)

IV. CONCLUSIONS

This paper proposed an oscillation source location method to find the location of the driving force of a forced oscillation that involves resonance with natural oscillation modes. The method is based comparing the oscillation mode shape of the forced oscillation with that of the natural oscillation. The method shows high accuracy and convenience in implementation in actual oscillation source location problems.